\documentclass[draft,grl]{AGUTeX}
\newcommand{\planss}{Planetary Space Science}

 \usepackage{lineno}
 \usepackage{multirow}
\linenumbers*[1]

  \usepackage[normalem]{ulem}
  \usepackage[pdftex]{graphicx}
  \usepackage{multirow}
  \usepackage{amssymb}
  \usepackage{amsmath}
  \usepackage{lscape}
  \usepackage{graphics}
\usepackage{epic}
\usepackage{ulem}
\usepackage{color}
\def\wp{W$^+$}
\def\op{O$^+$}
\def\ohp{OH$^+$}
\def\htwop{H$_2$O$^+$}
\def\htop{H$_3$O$^+$}
\def\h2o{H$_2$O}
\def\nh2o{$n_{\mathrm{H_2O}}$}
\def\re{R$_\mathrm{E}$}
\def\rem{\mathrm{R_{E}}}

\def\ne{$n_\mathrm{e}$}
\def\neh{$n_\mathrm{eh}$}

\def\mrat{$\dot{M}_\mathrm{exch}/\dot{M}_\mathrm{ioz}$}
\def\mioz{$\dot{M}_\mathrm{ioz}$}

\def\mexch{$\dot{M}_\mathrm{exch}$}

\def\mdot{$\dot{M}$}
\def\ux{$u_\mathrm{x}$}
\def\uxm{u_\mathrm{x}}

\def\vcom{v_\mathrm{cor}}
\def\vamm{v_\mathrm{amb}}

\def\vam{$v_\mathrm{amb}$}

\def\kgs{kg s$^{-1}$}

\providecommand{\fref}[1]{Figure \ref{#1}}
\providecommand{\sref}[1]{Section \ref{#1}}
\providecommand{\tref}[1]{Table \ref{#1}}
 \setkeys{Gin}{draft=false}
\authorrunninghead{FLESHMAN ET AL.}

\titlerunninghead{MODELING THE ENCELADUS PLUME--PLASMA INTERACTION}

\authoraddr{Bobby Fleshman,
Department of Physics and Astronomy, University of Oklahoma, 
440 W. Brooks, Norman, Oklahoma 73019, USA
(fleshman@nhn.ou.edu)}

\begin{document}

%
%

\title{Modeling the Enceladus Plume--Plasma Interaction}
%

%
%


\author{B. L. Fleshman}
\affil{Laboratory for Atmospheric and Space Physics,
University of Colorado, Boulder, Colorado, USA\\
Department of Physics and Astronomy, University of Oklahoma, 
Norman, Oklahoma, USA}
\author{P. A. Delamere}
\affil{Laboratory for Atmospheric and Space Physics,
University of Colorado, Boulder, Colorado, USA}
\author{F. Bagenal}
\affil{Laboratory for Atmospheric and Space Physics,
University of Colorado, Boulder, Colorado, USA}



%
%
%

%
\begin{abstract}
We investigate the chemical interaction between Saturn's corotating plasma and Enceladus' volcanic plumes.  We evolve a parcel of ambient plasma as it passes through a prescribed \h2o\ plume using a physical chemistry model adapted for water-group reactions.  The flow field is assumed to be that of a plasma around an electrically-conducting obstacle centered on Enceladus and aligned with Saturn's magnetic field, consistent with Cassini magnetometer data. We explore the effects on the physical chemistry due to: (1) a small population of hot electrons; (2) a plasma flow decelerated in response to the pickup of fresh ions; (3) the source rate of neutral \h2o.  The model confirms that charge exchange dominates the local chemistry and that \htop\ dominates the water-group composition downstream of the Enceladus plumes.  We also find that the amount of fresh pickup ions depends heavily on both the neutral source strength and on the presence of a persistent population of hot electrons.
\end{abstract}
%


%
%

%

\begin{article}
\section{Introduction}
Early Cassini encounters with Enceladus revealed surprising evidence of a significant source of water (with trace percentages of other neutrals, including CO$_2$) from geysers located at the moon's southern pole \citep{hansen2006, porco2006,spencer2006,waite2006}.  The \h2o\ cloud reacts with Saturn's corotating plasma torus, loading Saturn's magnetosphere with fresh ions.  The pickup rate \mdot\ quantifies the amount of fresh ions added to the magnetosphere from charge exchange and impact/photoionization.  Respective contributions to pickup from charge exchange and impact/photoionization are fundamentally different in that charge exchange does not contribute to ion production because one ion replaces another.  Both processes however introduce slow-moving ions which must subsequently be accelerated by Saturn's magnetosphere.

Early identification of the interaction between the local water source and Saturn's corotating plasma was made by \cite{dougherty2006}.  Based on the Cassini Plasma Spectrometer (CAPS, \cite{young2004}) analysis by \cite{tokar2006}, \cite{pontius2006} modeled the interaction and derived a pickup rate of \mdot\ $\approx$ 100 kg s$^{-1}$.  \cite{khurana2007} and \cite{saur2008} (hereafter K07 and S08) discovered a range in \mdot\ (0.2--3 kg s$^{-1}$) from Cassini magnetometer data in the three earliest Cassini Enceladus flybys (E0, 17 February 2005; E1, 09 March, 2005; E2, 14 July 2005).  Constrained by Ion Neutral Mass Spectrometer (INMS, \cite{waite2006}) and Ultraviolet Imaging Spectrograph (UVIS, \cite{hansen2006}) observations, \cite{burger2007} estimated a pickup rate of \mdot\ $\approx$ 2--3 kg s$^{-1}$ from a neutral cloud model.  The large discrepancy between the pickup rates derived from CAPS and magnetometer data is due not only to the fact that the region considered by \cite{pontius2006} is much larger than that considered by K07 and S08, but also because the \cite{pontius2006} result depends on the poorly-constrained value of Saturn's Pederson conductivity. 

In this paper, we use a physical chemistry model to investigate the chemical interaction between the corotating plasma and the Enceladus plumes.  Charge exchange dominates the local chemistry and leads to an \htop-dominated plasma downstream of Enceladus.  We find that pickup increases when hot electrons are present---more so with a high neutral source rate.

\section{Model}\label{sec:model}
We use a physical chemistry model developed to study the Enceladus torus \citep{delamere2003,delamere2007,fleshman2009} to investigate the composition of plasma traveling along prescribed flow lines.  The model evaluates mass and energy rate equations for water-group ions ($\mathrm{W}^+ \equiv$ \op + \ohp + \htwop + \htop), protons, and thermal electrons in a parcel of plasma transiting the simulation.  Neutrals are assumed to be cold, and in this study neutral abundances are fixed.  The full set of reactions includes charge exchange, photoionization, ionization by electron impact, radiative excitation, recombination, and molecular dissociation by both electron impact and recombination.  All species have isotropic Maxwellian speed distributions, and energy is transferred between species \textit{via} Coulomb collisions.  The simulation spans a rectangular domain extending 5\,\re\ from Enceladus in all directions except south, where the simulation extends to 15\,\re\ (\re\ $=252$ km is the radius of Enceladus).  

A second population of supra-thermal `hot' electrons is imposed with a fixed density (0.3 cm$^{-3}$) and temperature (160 eV).  Hot electrons near Enceladus have been reported by \cite{tokar2009} and have been observed throughout the torus by CAPS and the Radio and Plasma Wave Science Instrument \citep{moncuquet2005,young2005}.  We showed in \cite{fleshman2009} that a small amount of hot electrons is necessary to obtain the ambient ionization.  Here we investigate the importance of hot electrons near Enceladus itself.

\paragraph{Neutral source} Following S08, the plume is prescribed as
\begin{linenomath*}  
\begin{equation}
\label{eqn:nh2o}
\frac{n_\mathrm{H_2O}(r,\theta)}{n_0}=\left(\frac{\rem}{r}\right)^2\exp\left[-\left(\frac{\theta}{H_\theta}\right)^2-\left(\frac{r-\rem}{H_d}\right)\right],
\end{equation}
\end{linenomath*}  
where $H_\theta = 12^\circ$ and $H_d=948$ km (4 $\times$ the Hill radius).  S08 offset the plume from Enceladus' southern pole by 8$^\circ$ and considered more than one source with the form of (\ref{eqn:nh2o}).  We consider a single source whose origin coincides with Enceladus' south pole.  In the nominal case, $n_0$ is set to $2.5\times10^{9}$ cm$^{-3}$, corresponding to a neutral source rate of $\approx 200$ kg s$^{-1}$ (S08).  S08 found a much stronger source for E0, so we also investigate the implications of a source with $n_0=2.2\times10^{10}$ cm$^{-3}$, corresponding to a neutral  source rate of $\approx 1600$ kg s$^{-1}$.

\paragraph{Plasma flow field} Because of the low Alfv\'en Mach number at Enceladus ($M_\mathrm{A}$ $\approx$ 0.1, \cite{sittler2008}), perturbations travel rapidly along the magnetic field so that the source region presents a cylindrical obstacle to the corotating plasma.  We adopt the flow field used by \cite{dols2008} to study the plasma interaction with Jupiter's moon Io:
\begin{linenomath*}  
\begin{equation}
\label{u}
\frac{\mathbf{u}}{\vamm}= \left[1-\frac{\cos(2\phi)}{(\rho/\rem)^2} \right]\hat{\mathbf{x}}
-\frac{\sin(2\phi)}{(\rho/\rem)^2}\hat{\mathbf{y}},
\end{equation}
\end{linenomath*}  
where $\vamm\approx 0.8\times\vcom$ ($\vcom\approx$ 26 km s$^{-1}$) is the ambient plasma speed far from Enceladus \citep{wilson2009}.  The magnetic field defines the $z$-axis and $\phi$ is measured from the flow direction.

Along each flow line, parcels of plasma were started 5\,\re\ upstream of Enceladus with the steady-state composition given in \cite{fleshman2009}.  The plasma was moved in the direction of the plasma flow to 5\,\re\ downstream, and the chemistry was updated at associated time steps.  The pickup energy---determined by the relative speed between neutrals and plasma flow---was also updated.  Ions far from Enceladus are picked up at \vam, while those near Enceladus are picked up more slowly up- and downstream and more rapidly on the flanks.

To investigate the effect of ion pickup, we slowed the flow near the obstacle.  We followed \cite{dols2008} by decreasing the component of the plasma velocity in the flow direction:  \ux\ was replaced by $\gamma(\rho)\uxm$, where $\gamma(1\,\mathrm{R_E})=0.5$, increasing linearly to $\gamma(2\,\mathrm{R_E})=1$.  We find however that a stronger slowing factor ($\gamma(1\,\mathrm{R_E})=0.1$) does not qualitatively change our results.  In \sref{section:results}, we compare the pickup rates both for when the plasma has (nominal case) and has not been slowed.

Two effects are due directly to the slowing of the flow.  First, the pickup energy is reduced, affecting the plasma temperature because fresh ions are picked up at the local plasma speed.  Second, impact ionization increases because plasma transits more slowly.  Impact ionization contributes directly to pickup, as well as indirectly, by seeding multiple charge exchanges.

We neglect gyromotion on the basis of scale.  For example, an \htop\ ion picked up at \vam\ has a gyroradius of only 0.1\,\re.  More important is that (in the frame of Enceladus) ions oscillate between zero and twice their pickup speed and we ignore the velocity dependence of charge exchange.  Including gyromotion would enhance the effects we report in this paper.

\paragraph{Hot electrons} We estimate that hot electrons cool rapidly near the dense plume \textit{via} impact ionization, and thus imposed a discontinuity at $\rho=3$\,\re: 
\begin{linenomath*}  
\begin{equation}
\label{eqn:hotElectrons}
n_\mathrm{eh}/(\text{cm}^{-3}) = 
\left\{ 
\begin{array}{lll} 
0  &1&<\rho/\mathrm{R_E}<3\\
 0.3 \ \ \ \  &3&<\rho/\mathrm{R_E}. 
\end{array} 
\right. 
\end{equation}
\end{linenomath*}  

However, \cite{dols2008} showed that field-aligned electron beams (perhaps associated with the Io auroral footprint) are necessary to model the high plasma density in Io's wake.  At Enceladus, hot electrons may also be related to weak UV auroral spots recently observed by UVIS (W. Pryor, personal comm.).  To investigate the implication of hot electrons at Enceladus, we consider the additional case where \neh\ is held at 0.3 cm$^{-3}$ throughout the simulation domain.  The pickup rates for each case are compared in \sref{section:results}.

\paragraph{Pickup rate calculation} Fresh ions are added to the magnetosphere by both charge exchange ($\dot{\rho}_{\mathrm{exch}}$) and impact/photoionization ($\dot{\rho}_{\mathrm{ioz}}$):
\begin{linenomath*}  
\begin{equation}
\label{rhoioz0}
\dot{\rho}_{\mathrm{ioz}} = \sum_j n_\mathrm{e} n_j m_j \kappa_j^{\mathrm{imp}} + \sum_k n_k m_k \kappa_k^{\mathrm{phot}}
\end{equation}
\begin{equation}
\label{rhoexch0}
\dot{\rho}_{\mathrm{exch}} = \sum_j n^{(1)}_j n^{(2)}_j m_j \kappa_j^{\mathrm{exch}}.
\end{equation}
\end{linenomath*}  
The reaction rates (\cite{fleshman2009}) are represented by $\kappa^{\mathrm{imp}}$, $\kappa^{\mathrm{exch}}$ [cm $^3$ s$^{-1}$], and $\kappa^{\mathrm{phot}}$ [s$^{-1}$]; the ion masses by $m$; and $n^{(1)}$, $n^{(2)}$ are the charge-exchanging neutral and ion densities.  Summations are carried out over processes involving the creation of fresh ions.

We calculated the time-averages of (\ref{rhoioz0}) and (\ref{rhoexch0}) to find the average pickup rates for plasma parcels migrating along each flow line and multiplied by the flow line volume to find the pickup rate for each flow line.  The total pickup rate for each process was determined by summing the contribution from all flow lines throughout the simulation.

\section{Results} \label{section:results}
\begin{figure}
\noindent\includegraphics[width=6.5in,angle=0]{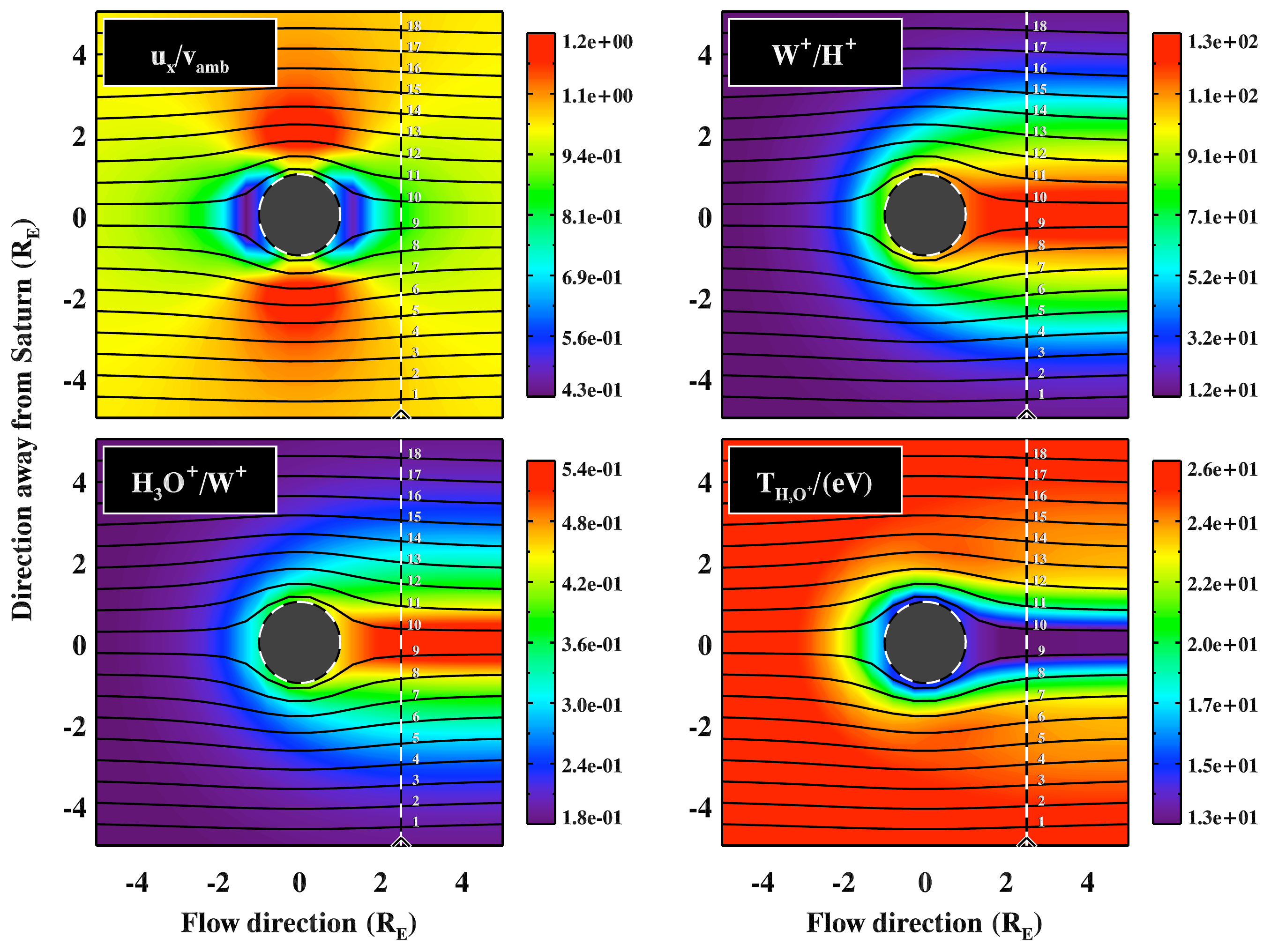}
 \caption{\textit{Upper-left}:  the $x$ component of the assumed local plasma flow speed (\sref{sec:model}) normalized to the ambient flow speed (80\% of rigid corotation). \textit{Upper-right}:  water-group to proton abundance ratio.  \textit{Lower-left}:  \htop/W$^+$ abundance ratio.  \textit{Lower-right}:  \htop\ temperature.  The plane represented here is 7.5\,\re\ south of Enceladus.  Model output along the dashed line is given in \fref{fig3}.} 
 \label{fig2}
\end{figure}

\begin{figure}
\noindent\includegraphics[width=3.5in,angle=0]{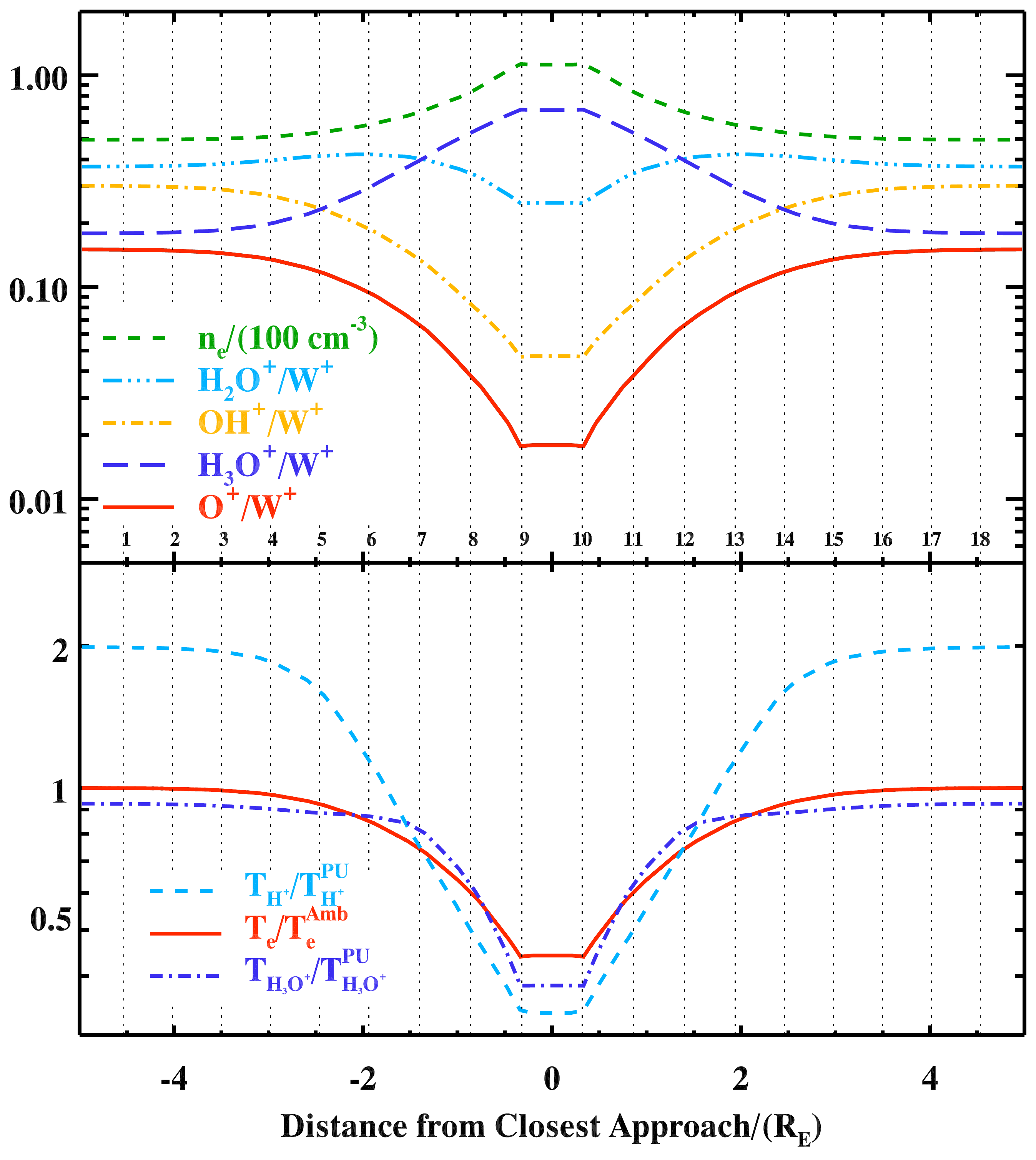}
 \caption{\textit{Top}:  Abundances and electron density from the simulation along the dashed line in \fref{fig2}.  \textit{Bottom}:  Electron, proton and \htop\ temperatures along the same cut.  The electron temperature is normalized to its ambient temperature (2 eV), and the ion temperatures are normalized to their ambient pickup energies---1.5 and 29 eV for H$^+$ and H$_3$O$^+$ respectively.}
 \label{fig3}
\end{figure}

We consider the $x$--$y$ plane 7.5\,\re\ south of the center of Enceladus.  For this, the nominal case, hot electrons exist throughout the domain, the neutral source rate is $\approx$\,200 kg s$^{-1}$, and the plasma is slowed in the flow direction (\sref{sec:model}).  The pickup energy, heavy-to-light ion abundance (W$^+$/H$^+$), \htop/W$^+$, and \htop\ temperature ($T_\mathrm{H_3O^+}$) in this plane are shown in \fref{fig2}.  The obstacle is plotted in black and 18 flow lines are over-plotted and labeled.  Composition and temperatures along this line are plotted in \fref{fig3}, where the corresponding flow lines are indicated.

The \htop\ ion is the most abundant ion in the wake---a consequence explained by the importance of charge exchange and the fact that charge exchanges lead to \htop\ in the presence of an abundant water source.  All charge exchanges with \h2o\ in our model ultimately lead to either \htop\ or \htwop\ by $\mathrm{H_2O^+ +H_2O \rightarrow H_2O +H_2O^+}$ and $\mathrm{H_2O^+ +H_2O \rightarrow OH +H_3O^+}$.  The former reaction supports \htwop\ density somewhat, but the \htwop\ products also feed into the latter, producing \htop.  The $\mathrm{W^+/H^+}$ ratio increases rapidly because protons are removed by $\mathrm{H^+ +H_2O \rightarrow H +H_2O^+}$.  The increase in electron density is due mainly to impact ionization of the plume by hot electrons.

In \fref{fig3}, the electron temperature has been normalized to the ambient electron temperature (2 eV), and the ion temperatures have been normalized to their respective ambient pickup energies (1.5 and 29 eV).  Protons and \htop\ bear the signature of the cooler pickup temperature from where they were created by charge exchange.  A factor of a few decrease in ion temperature through Enceladus' wake has recently been observed by \cite{tokar2009}.  The electron temperature has also been cooled by a factor of 2.

\paragraph{Pickup rate} We calculated individually (\sref{sec:model}) the mass of fresh ions picked up \textit{via} charge exchange, \mexch, and \mioz$-$the ratio of which illustrates the importance of charge exchange over impact ionization.  The simulation was run for the eight cases shown in \tref{table:massLoad}.  The flow-field, hot-electron, and source-rate treatments are described in \sref{sec:model}.  We discuss the cases corresponding to a `weak' 200 \kgs\ source (Cases 1a--4a) and a `strong' 1600 \kgs\ source (Cases 1b--4b) separately.

\textit{Weak source (Cases 1a--4a):} When hot electrons exist locally (1a/2a), the total pickup is roughly 0.3 \kgs\ with a 40\% increase when the plasma is slowed.  Because of the longer occupation time associated with the slowed flow, hot electrons increase seed ionization (\mioz), and in turn increase charge exchange (\mexch).  When hot electrons are removed (3a/4a), \mioz\ decreases by a factor of 4, but \mexch\ drops by $\approx$\,30\%, implying that much of the pickup is occurring outside the cut-off point at 3\,\re.  Slowing the plasma has less effect in 3a than in 1a because longer occupation does not boost seed ionization without hot electrons.

\textit{Strong source (Cases 1b--4b):} The effect of the hot electrons becomes more apparent with the strong source because a denser portion of the plume is intersected by the flow lines.  The effect of slowing the plasma is similar to that in the weak-source case.  The total pickup however is increased by a factor of 3 when comparing the cases with hot electrons (1b/2b) to cases without (3b/4b).  The almost linear response of the total pickup to \mioz\ (compare 1b/2b to 3b/4b) suggests that, unlike in the weak source case, most of the pickup is occurring inside the hot electron cutoff at 3\,\re.  
 
K07 and S08 were in rough agreement on the total pickup rate.  For E1 and E2, they found $\approx 0.2$--$0.6$ kg s$^{-1}$, and for E0 they found $\approx 3$ kg s$^{-1}$.  Because our model relies on a physical chemistry calculation alone, it is remarkable to have obtained the same pickup rates using neutral plume distributions similar to those in S08 (E0: strong source, E1 and E2: weak source).  

Charge exchange dominates the chemistry by at least a factor of 6 in all cases.  Because of this, the water-group composition ratios (shown only for the nominal case) are qualitatively unaffected.  In particular, \htop/\wp\ always increases while \op/\wp, \ohp/\wp, and \htwop/\wp\ always decrease in Enceladus' wake.  The dominance of \htop\ elsewhere has been observed by CAPS during Cassini's orbital insertion period \citep{tokar2006,sittler2008}.  \cite{fleshman2009} found that a steady-state, water-based Enceladus torus underestimates the \htop\ abundance seen in the CAPS data.  Though few \textit{new} ions are produced at Enceladus, the process that produces \htop\ may have an important effect on the large-scale torus composition.  A more complete global model of the torus should include the effect of dense \h2o\ on Saturn's corotating plasma demonstrated in this paper.

\section{Discussion and conclusions}
To investigate the impact of hot electrons on the chemistry, we have chosen the simplest flow field possible with minimal perturbation by Enceladus.  This flow is roughly consistent with the compact source derived from magnetometer data (K07) but is much less perturbed than the flow reported by \cite{tokar2006}.  Similarly, we have started with a single symmetric plume oriented due-south.  In future studies we will explore multiple jets (S08), displaced sources (K07), and a minor spherically-symmetric global component \citep{burger2007}.

Our findings are summarized below:
\begin{enumerate}
  \item Charge exchange dominates the plume--plasma chemistry, confirming previous work by \cite{burger2007} and consistent with estimates by \cite{johnson2006}.
  \item Charge exchange leads largely to an \htop-dominated wake, consistent with INMS \citep{cravens2009}.  Reactions leading to \htop\ are well known in the comet community \citep{aikin1974,haberli1997}.
  \item Comparing our pickup rates to those derived from the Cassini magnetometer (K07 and S08), INMS, and UVIS \citep{burger2007}, we find that a persistent source of hot electrons may exist near Enceladus.  If present, beams of hot electrons at Enceladus may be related to the weak UV auroral spots recently observed by W. Pryor (personal comm.).
\end{enumerate}

%
%

%
%
%
%
%
%

%
%
%
%
\begin{acknowledgments}
This work was supported under NESSF grant Planet09F--0036 and CDAP grant 06-CASS06-0062.
\end{acknowledgments}

%
%
%
%
%
%
%
%
%
%


%
%

\end{article}




%
%
%
%
%
%




%
\newpage
\clearpage
\pagebreak
\begin{table}
\centering      
\begin{tabular}{l c @{\ $+$\ } c| r@{.}l  r@{.}l | r@{.}l r@{.}l |c}  
\hline\hline                         
& \multicolumn{2}{c}{\rule{0pt}{2.6ex}\multirow{2}{*}{Case}} 
& \multicolumn{4}{c}{\mexch} 
& \multicolumn{4}{c}{\mioz} 
& \multirow{2}{*}{\mrat} \\ 
& \multicolumn{2}{c}{\ } 
& \multicolumn{2}{c}{\ \ kg s$^{-1}\ $} 
& \multicolumn{2}{c}{$(10^{25}\ \mathrm{H_2O\ s^{-1}})$}
& \multicolumn{2}{c}{\ \ kg s$^{-1}\ $} 
& \multicolumn{2}{c}{$(10^{25}\ \mathrm{H_2O\ s^{-1}})$}
& \  \\ [0.2ex] 
\hline                    
\rule{0pt}{2.6ex}\textbf{(1a)} & \textbf{Hot electrons} & \textbf{slowed flow}
&\ \  \textbf{0}&\textbf{25}    &\ \ \ \ \textbf{(0}&\textbf{84)}
&\ \  \textbf{0}&\textbf{038}  &\ \ \ \ \textbf{(0}&\textbf{13)}
& \textbf{6.7}  \\ 
(2a) & Hot electrons & un-slowed flow\ \ 
& 0&22    & (0&74)
& 0&034  & (0&11)
& 6.6  \\ 
(3a) & No hot electrons & slowed flow
& 0&18    & (0&61)
& 0&0091  & (0&030)
& 20  \\ 
(4a) & No hot electrons & un-slowed flow
& 0&17   & (0&58)
& 0&0087   & (0&029)
& 20  \\ 
\hline     
\rule{0pt}{2.6ex}(1b) & Hot electrons & slowed flow
& 2&2    & (7&3)
& 0&33    & (1&1)
& 7.0  \\ 
(2b) & Hot electrons & un-slowed flow
& 1&9    & (6&5)
& 0&30  & (0&99)
& 6.6  \\ 
(3b) & No hot electrons & slowed flow
& 0&77    & (2&6)
& 0&079  & (0&26)
& 9.8  \\ 
(4b) & No hot electrons & un-slowed flow 
& 0&77    & (2&6)
& 0&076  & (0&25)
& 10  \\ 
\hline     
\end{tabular} 
\caption{Pickup rates from charge exchange (\mexch) and impact/photoionization (\mioz) for the eight cases discussed in the text.  Only \mioz\ increases the plasma density (\ne).  The cases labeled with `a' correspond to a neutral source rate of 200 kg s$^{-1}$;  those labeled with `b' correspond to a neutral source rate of 1600 kg s$^{-1}$.  Case (1a), in bold, is the nominal case from which Figures \ref{fig2} and \ref{fig3} have been generated.} 
\label{table:massLoad}  
\end{table} 
\pagebreak
\pagebreak
\end{document}